\documentclass{article} 
\usepackage{iclr2021_conference,times}

\usepackage{amsmath,amsfonts,bm}

\def\eqref#1{equation~\ref{#1}}

\def\1{\bm{1}}

\DeclareMathAlphabet{\mathsfit}{\encodingdefault}{\sfdefault}{m}{sl}
\SetMathAlphabet{\mathsfit}{bold}{\encodingdefault}{\sfdefault}{bx}{n}

\def\gT{{\mathcal{T}}}

\DeclareMathOperator*{\argmax}{arg\,max}

\newcommand{\eq}[1]{\begin{align}#1\end{align}}

\newcommand{\brck}[1]{\left(#1\right)}

\newcommand{\brcksq}[1]{\left[#1\right]}
\newcommand{\brckcur}[1]{\left\{#1\right\}}

\usepackage{hyperref}
\usepackage{url}
\usepackage{graphicx}
\usepackage{caption}
\usepackage{subcaption}

\def\ourframework{RIRL}
\def\ourpolicy{RIRL-actor}

\def\timev{v}
\def\timet{t}

\newcommand{\policy}{\pi}

\def\calI{\mathcal{I}}

\def\ob{o}

\def\ac{a}

\def\st{s}

\def\rew{r}
\def\Rew{R}

\def\df{\gamma}

\def\trans{\gT}

\def\lstmstate{h}
\def\discriminator{D}
\def\decoder{\omega}
\def\encoder{q}

\def\MI{I}
\def\MCMI{\tilde{I}} 
\newcommand{\MCMIgen}[1]{\MCMI_{#1}}
\def\mimult{\lambda}
\newcommand{\RI}[1]{#1^{\dagger}}

\def\utility{u}
\def\horizon{T}
\def\Chn{M} 
\def\chn{m} 
\def\Type{K} 

\def\hours{h} 
\def\effort{e} 
\def\labor{l} 
\def\output{z}
\def\Output{\mathcal{Z}}
\def\ability{\nu}
\def\wage{w} 
\def\numa{n_a} 
\def\nagents{N}

\def\payment{w}
\def\schedule{\mathcal{W}}

\def\aplamb{\lambda_{a_p}} 
\def\elamb{\mimult^\effort} 
\def\olamb{\mimult^\output} 

\newcommand{\parwrapper}[1]{\paragraph{#1}}

\title{Solving Dynamic Principal-Agent Problems with a Rationally Inattentive Principal}
 \author{Tong Mu\textsuperscript{†} \thanks{Work done while an intern at Salesforce Research.} \\
 Stanford University
 \And
 Stephan Zheng \thanks{tongm@stanford.edu, \{stephan.zheng, atrott\}@salesforce.com}\\
 Salesforce Research
 \And
 Alexander Trott\textsuperscript{†}\\
 Salesforce Research
 }

\iclrfinalcopy 
\begin{document}

\maketitle

\begin{abstract}
Principal-Agent (PA) problems describe a broad class of economic relationships characterized by misaligned incentives and asymmetric information.
The Principal's problem is to find optimal incentives given the available information, e.g., a manager setting optimal wages for its employees.
Whereas the Principal is often assumed rational, comparatively little is known about solutions when the Principal is boundedly rational, especially in the sequential setting, with multiple Agents, and with multiple information channels.
Here, we develop \ourframework{}, a deep reinforcement learning framework that solves such complex PA problems with a \emph{rationally inattentive} Principal. 
Such a Principal incurs a cost for paying attention to information, which can model forms of bounded rationality.
We use \ourframework{} to analyze rich economic phenomena in manager-employee relationships.
In the single-step setting, 
1) \ourframework{} yields wages that are consistent with theoretical predictions; and
2) non-zero attention costs lead to simpler but less profitable wage structures, and increased Agent welfare.
In a sequential setting with multiple Agents, \ourframework{} shows opposing consequences of the Principal's inattention to different information channels:
1) inattention to Agents' outputs closes wage gaps based on ability differences; and
2) inattention to Agents' efforts induces a social dilemma dynamic in which Agents work harder, but essentially for free.
Moreover, \ourframework{} reveals non-trivial relationships between the Principal's inattention and Agent types, e.g., if Agents are prone to sub-optimal effort choices, payment schedules are more sensitive to the Principal's attention cost.
As such, \ourframework{} can reveal novel economic relationships and enables progress towards understanding the effects of bounded rationality in dynamic settings.\footnote{Example code is available at \url{https://github.com/salesforce/RIRL}}
\end{abstract}
\section{Introduction}

\paragraph{Principal-Agents Problems.}
Principal-Agent (PA) problems are well-established in the economics literature and occur in many different contexts where multiple parties face misaligned incentives and asymmetric information.
A salient application is the design of payment contracts between a manager and its employees.
Broadly speaking, the Principal's problem is to find optimal incentives, subject to the limited information it can use.
For example, as a common modeling assumption, a manager (Principal) may not see the skill or amount of effort exerted by an employee (Agent). 
Hence, employee payment can be contingent on the employee's output perceivable by the manager, but this may correlate imperfectly with skill or the actual effort exerted. 
As an example failure mode, a project may fail due to insufficient employee effort or due to circumstances beyond an employee's control, even though enough effort was exerted.

Prior work has examined how risk aversion limits efficiency in PA problems~\citep{Holmstrom1987, Spremann1987, haubrich1994risk}.
A related line of work considers cases where the Principal must infer wage-relevant attributes of the Agent and the incentive structures embedded in such dynamics~\citep{Spence1973, Alos-Ferrer2012, holmstrom1999managerial}.
In both settings, information asymmetry is key to the problem formulation.

\paragraph{Understanding the Effect of Bounded Rationality.}
Whereas typical analyses solve the Principal's problem assuming rationality, comparatively little is known about solutions when the Principal is \textit{boundedly rational}.
This is a significant gap in the literature, as it is unrealistic to assume that stylized, rational Principals can fully characterize real-world, human Principals.
Indeed, laboratory PA experiments with human participants have shown that bounded rationality is needed to explaining marked deviations between equilibria reached by human participants and theoretical predictions \citep{erlei2017bounded}.

In this work, we explore the implications of bounded rationality in PA problems by modeling a Principal with \textit{inattention}.
Inattention helps to explain a broad range of human behavior \cite{Gabaix2017}. 
The basic intuition for the relevance of inattention is that cognitive limitations restrict the amount of information an actor can consider when making a decision.

In the context of PA problems, this suggests that a Principal's inattention constitutes an \emph{endogenous} source of information asymmetry.
In other words, bounded rationality is implemented by forcing the Principal to \emph{choose} which information it will use to make decisions, by introducing a \emph{cost to paying attention}.

Theoretically analyzing such inattention effects quickly becomes intractable, especially so in the sequential setting, with multiple Agents, and with multiple information channels.
For one, incorporating attention introduces an additional, complex decision space.
Moreover, the marginal value of paying attention depends non-trivially on both its cost and how better-informed Principal decisions change the Agents' incentives and behavior.

The difficulty of these technical challenges is reflected in a substantial lack of research progress in this direction.
One noteworthy exception is found in~\citet{mirrlees1976optimal}, which considered payment contracts when it is costly for the Principal to accurately assess output.
However, the resulting insights were mostly qualitative and limited by difficulties in advancing the analytical treatment.

\paragraph{Solving PA Problems through \ourframework{}}
To enable progress in the face of these challenges, we introduce \ourframework{}: a computational framework for solving and analyzing complex, dynamic PA problems with an inattentive Principal.
\ourframework{} uses multi-agent reinforcement learning (MARL), a tool that has been shown to be fruitful for analyzing economic behavior in rich economic environments~\citep{zheng2020ai,baker2019emergent,Leibo2017,Yang2018}.
This framework optimizes both the Principal's and Agents' behaviors to maximize their utility, this including a cost for attention for the Principal.
These costs are defined following Rational Inattention (RI), a well-established model of bounded rationality that formalizes the cost of attention as the mutual information between (1) variables relevant to the decision and (2) the decision itself~\citep{sims2003implications,Mackowiak2011,mackowiak2021rational}.

\paragraph{Contributions and Findings.}
In summary, our work contributes
\begin{enumerate}
    \item an effective methodological framework for analyzing rational inattention in PA problems, 
    \item a demonstration of soundness by showing \ourframework{} aligns with classic theory for the problem of setting payment schedules, as well as
    \item a detailed economic analysis of the effects of inattention in both a single-step and dynamic settings with multiple Agents and information channels.
\end{enumerate}
In particular, \ourframework{} can reveal and quantify non-trivial economic relationships, especially in generalized PA problems that are analytically intractable, e.g., a sequential PA problem with multiple Agents and heterogeneous information channels.

For example, in a single-step setting, \ourframework{} reveals non-trivial relationships between the Principal's inattention and Agent types, e.g., if Agents are prone to sub-optimal effort choices, payment schedules are more sensitive to the Principal's attention cost.
In a sequential setting with multiple Agents, \ourframework{} shows that information channels can have opposing consequences, e.g., closing wage gaps or creating social dilemmas that induce unrewarded effort.

As such, \ourframework{} enables progress towards understanding the effects of bounded rationality in dynamic settings.
\section{Related Works}
\label{section:related-works}


\parwrapper{Principal-Agent Problems.}
Principal-Agent (PA) problems have been used to describe various settings, including politics and insurance~\citep{miller2005political,grossman1992analysis}. 
Prior economics literature mainly use analytical methods and narrow modeling assumptions, e.g., that all stochasticity follows Brownian motion~\citep{sannikov2008continuous} or certain separability conditions are satisfied ~\citep{grossman1992analysis}.
Generally, Principal-Agent settings fall under the category of Stackelberg games where it has been shown that computing optimal strategies can be NP-hard~\citep{korzhyk2010complexity}.
Instead, multi-agent RL (MARL) is a constructive solution for complex setups that are analytically intractable.
\citet{shu2018m,shi2019learning,ahilan2019feudal} studied coordinating cooperation in a Principal-Agent model, with an RL Principal learning to incentivize Agents to achieve an overall goal.
However, they do not consider bounded rationality.

\parwrapper{Models of Bounded Rationality.}
Behavioral economics has shown extensively that human decision-making is not fully rational but instead features many cognitive biases~\citep{caverni1990cognitive}. 
Bounded rationality~\citep{simon1957models} attributes these human irrationalities to resource limitations, e.g., bounded cognitive capabilities or costliness of using (more) time to make decisions. 
Rational Inattention (RI) is a well-established model of bounded rationality that explains irrational behavior as ``rational'' when optimizing for an objective that includes an attention cost~\citep{sims2003implications,mackowiak2021rational}.
RI has been tested in real world experiments~\citep{dean2017experimental} and used to model human behavior in a wide variety of domains~\citep{hoiles2020rationally,mackowiak2021rational}. %
Similar to our work but in a different domain, \citet{jiang2019route} study a multi-agent system for traffic route choice under RI but where the agents do not react to each other's actions.
Another line of work uses deep learning to predict human behavior~\citep{bourgin2019cognitive,kolumbus2019neural,hartford2016deep} and model cognition~\citep{kubilius2019brain,battleday2017modeling,ma2020neural} from datasets of human behavior.
In contrast, our approach examines the implications of RI in simulated PA settings where experimental data are presently unavailable.

%

\parwrapper{Multi-Agent Reinforcement Learning.} 
Multi-Agent Reinforcement Learning (MARL)-based simulations use RL agents which autonomously learn utility-maximizing behavior, so designers do not need to specify behavioral rules.
MARL has been used to study tax policy~\citep{zheng2020ai,Zheng2021ai,Trott2021}, games~\citep{baker2019emergent}, and social dilemmas~\citep{Leibo2017} among others~\citep{Yang2018}. 
However, RL agents are mostly assumed rational which contradicts with human behavior.
Some recent work considers bounded rationality in MARL by accounting for cognitive limitations when reasoning about the behaviors of other agents~\citep{evans2021bounded,wen2019modelling,latek2009bounded}, which complements inattention as a source of bounded rationality.
In economics, Agent-Based Models (ABM) are a related line of work that study emergent phenomena through simulation with fixed or manually-specified agent behaviors~\citep{Bonabeau2002}, and, for example, to what extent those phenomena resemble real-world systems.

\parwrapper{Mutual Information in Reinforcement Learning.}
MI has been extensively studied in the RL literature for curiosity-driven exploration and unsupervised skill and option discovery~\citep{still2012information,campos2020explore,mohamed2015variational,gregor2016variational,eysenbach2018diversity}. 
These works often use differing techniques to measure MI.
MI has also been used to regularize exploration~\citep{grau2018soft, leibfried2020mutual}. 
Comparatively little work has considered MI-based rewards for modeling boundedly rational behavior~\citep{Ortega2015, peng2017information}, with no prior work, to our knowledge, considering PA problems.
\section{Formal Definitions and Problem Statement}
\label{section:preliminaries}

To define our problem of interest, we first introduce the Principal-Agent problem, assuming all actors are rational. 
We then introduce a Principal who is rationally inattentive and outline what challenges we address in this setting.

\subsection{Principal-Agent Problems}

Principal-Agent (PA) problems study how a Principal (e.g., a manager) can align the incentives and behaviors of (a set of) Agents (e.g., employees) with its own.
A key feature is information asymmetry: the Principal often cannot observe the utility function, exerted effort, or other private information of the Agents.
As such, the Principal needs to optimize its policy (e.g., setting wages) based on limited knowledge and anticipating that the utility-maximizing behavior of the Agents may not improve the Principal's objective.

We model Principal-Agent problems with 1 Principal and $\nagents$ Agents as partially-observable Markov Games (MGs). 
An MG is formally defined as a tuple $(S, A, \rew, \trans, \df, O, \calI)$ \citep{sutton_reinforcement_2018}.
Here $S$ is the state space of the game, $A$ is the combined action spaces of the actors, and $\calI$ are actor indices, where the actors are the Principal and the $\nagents$ Agent(s).
The Principal is labeled by $i=p$, while Agents are labeled by $i=1, \ldots, \nagents$.

We use $\ob_i = O(s, i)$ to denote the portion of the game state $s \in S$ that actor $i \in \calI$ can observe.
In addition, $\ob_i$ may include a (possibly learnable) encoding of the observation history.
Each game episode has a horizon of $\horizon \geq 1$ timestep(s).
Each timestep $t$, actor $i$ selects action $a_{i,t}$, sampling from its policy $\policy_i \left( a_{i,t} | \ob_{i,t} \right)$.
Given the sampled actions, the transition function $\trans$ determines how the state evolves.\footnote{Actors do not necessarily act simultaneously within a single timestep; for example, we model timesteps where the Principal first takes its action (to set payment conditions) and then the Agent(s) act given knowledge of the Principal's choice.}
Each actor's objective is encoded in its reward function $\rew_i(s, \bm{a})$, where boldface denotes concatenation across actors.
When modeling economic behavior, the reward is the utility $\utility_i(s_t, \bm{a}_t)$.
A rational actor optimizes its policy $\policy_i$ to maximize its $\df$-discounted return (sum of future rewards):
\eq{
\policy_i^* = \argmax_{\policy_i}\mathbb{E}_{\bm{\policy}, \trans} \brcksq{ \sum_t \df^t \rew_i(s, \bm{a}) }.
}

We use the subscript notation $\ac_{-i}$ to indicate a vector with elements that are \emph{not} pertinent to actor $i$.
The solution to the PA problem is then to find the optimal Principal policy:
\eq{
\label{eq:pa-problem-principal}
\policy_p^* &= \argmax_{\policy_p} \mathbb{E}_{\policy_p, \bm{\policy}^*_{-p}, \trans} \brcksq{\sum_{t}\df^t \utility_p(\st_t, \ac_{t;p}, \bm{\ac}^*_{t;-p})}, \\
\label{eq:pa-problem-agents}
& \text{s.t. } \ac^*_{\timet;i} \sim \policy_i^*\brck{.|\ob_\timet}, \quad 
\policy^*_i\brck{.|\ob_\timet} = \argmax_{\policy_i} \mathbb{E}_{\policy_p, \bm{\policy}_{-p}, \trans} \brcksq{\left. \sum_{\timev\geq \timet} \df^\timev \utility_i\brck{\st_\timev, \ac_{\timev;p}, \bm{\ac}_{\timev;-p}}  \right| \ac_{\timet;p} }, \quad 
\forall i \neq p. 
}
Equation \ref{eq:pa-problem-principal} states that the Principal optimizes its return assuming the Agents play a best response to the Principal, i.e., use a policy $\policy_i^*$ that maximizes their return.

\subsection{Modeling a Principal with Rational Inattention}
A common assumption in the PA problem is that the Principal is rational and can observe any available information \emph{at no cost}. 
However, it is well-known that human behavior is often not rational, as studied in behavioral economics. 
In particular, \textit{inattention} has been broadly used as a mechanism that can explain irrational human behavior \cite{Gabaix2017}.

Inattention models the idea that paying attention to (sources of) information can be \emph{costly}.
For instance, it may take time and energy for a manager to observe the work performance of individual employees of a team, as opposed to just observing the team's output. 
Due to this attention cost, a rationally inattentive actor may or may not choose to pay attention to certain information.

Rational Inattention (RI) explains such behavior as optimal by adding the \textit{attention cost} to the actor's objective.
Specifically, RI uses a modified objective that includes a cost to the mutual information $\MI(\ac_i; \ob_i)$ between the (observable) state of the world $\ob_{i}$ and the actions $\ac_{i} \sim \policy_i(\cdot | \ob_{i})$. 
The Principal's objective becomes:
\eq{
\label{eq:rational-inattention-objective}
\RI{\policy}_p &= \argmax_{\policy_p} \mathbb{E}_{\policy_p, \bm{\policy}^*_{-p}} \brcksq{\sum_\timet \df^\timet \RI{\utility}_p(\st_t, \ob_{\timet;p}, \bm{a}_t; \mimult) }, \\
\RI{\utility}_{p}(\st_\timet, \ob_{\timet;p}, \bm{\ac}_\timet; \mimult) &= \utility_p(\st_\timet, \bm{\ac}_\timet) - \mimult \MCMI(\ac_{\timet;p}; \ob_{\timet; p}),
}
where $\mimult$ is the utility cost per bit of information (i.e. the \textit{strength} of the attention cost paid by the Principal).
Here the mutual information $\MI$ quantifies how much information the actor requires about the state of $\ob_i$ (i.e., the amount of attention it must pay) to execute its policy $\policy_i$:
\eq{\label{eq:mutual-info}
\MI(\ac_{\timet; p}; \ob_{\timet; p}) &= 
\log \frac{
p(\ac_{\timet; p}, \ob_{\timet; p})
}{
p(\ac_{\timet; p})p(\ob_{\timet; p})
}.
}
Here $p(\ac_p, \ob_p)$, $p(\ac_p)$, and $p(\ob_p)$ are the joint and marginal distributions over $\ac_p$ and $\ob_p$ induced by the environment and policies $\bm{\policy}$.
%

To build intuition for the use of $\MI$, imagine the extreme case where the actor pays \textit{no} attention to $\ob_i$.
The probability of a given action being selected wouldn't depend on the observation and the MI cost $\MI(\ac_i; \ob_i)$ would be 0.
On the other hand, if the action correlates with the observation, the Principal is assumed to have paid attention to $\ob_i$ in order to select $\ac_i$ and $\MI > 0$.

From a learning perspective, the MI cost acts as a regularizer which encourages simpler policies, i.e., where actions are less dependent on the precise state of the world and/or history of observations.
Rationally inattentive actors therefore are incentivized (so to speak) to select which information they use, making inattention an \emph{endogenous} source of information asymmetry.

From an economic perspective, the attention cost may model limits on information processing and hence restrict the types of incentives the Principal can provide.
Alternatively, attention costs could reflect actual expenditures, i.e., monitoring costs.

\subsection{Multiple Information Channels}
To model more general forms of rational inattention, we extend our framework to support 

\begin{enumerate}
    \item multiple channels of information with heterogeneous cognitive costs, and
    \item modeling which observation channel an agent pays attention to.
\end{enumerate}
For example, when buying a used car it is easier to see car prices (which might still take time) than ascertaining the actual condition of the car.

As such, we first extend the \emph{action-perception decoupling} strategy of~\citet{peng2017information} to model $\policy(\ac | \ob)$ as a stochastic perception module $\encoder(y | \ob)$ (encoder) followed by an action module $\decoder(\ac | y)$ (decoder). 
The RI utility then becomes: 
\eq{
\RI{\utility} = \utility(\st, \ac) - \mimult_{\encoder} \MI_{\encoder}(y; \ob) - \mimult_\decoder \MI_\decoder(\ac; y),
}
where we omitted subscripts for clarity.
This decomposes the attention cost into: (1) the cost of paying attention to the observation $\ob$, and (2) the cost of using that information to select an action.

Second, we extend the encoder to capture attention costs that depend on $\Chn$ parallel observations $\{\ob^1, \ldots, \ob^\Chn\}$, giving an RI utility function:
\eq{ 
\RI{\utility} = \utility(\st, \ac) - \mimult^1 \MI(y^1; \ob^1) \ldots - \mimult^\Chn \MI(y^\Chn; \ob^\Chn) -  \mimult_\decoder \MI_\decoder(\ac; \left[y^1, \ldots, y^\Chn \right]),
}
where brackets denote concatenation, and each channel has its own attention cost $\mimult^\chn$ of observing $\ob^\chn$.

\section{A Learning Framework for Rational Inattention}
\label{section:methods}

We now describe a practical approach to estimating the attention cost in Equation \ref{eq:rational-inattention-objective} and solving the optimization problem in Equations \ref{eq:pa-problem-principal} and \ref{eq:pa-problem-agents}, assuming the Principal's reward include rational inattention as specified in Equation \ref{eq:rational-inattention-objective}.

These optimization problems present several technical challenges:
\begin{itemize}
    \item Finding the Principal's optimal policy is challenging in a sequential setting and with multiple Agents.
    \item Computing the true mutual information cost $\MI$ is computationally expensive.
\end{itemize}

Our framework \ourframework{} solves the Principal's problem using reinforcement learning (RL), while we propose a sampling-based estimator for $\MI$ that works well in practice.

\subsection{Mutual Information Estimation}
\label{sec:mutual-info-estimation-single-obs}
The key challenge in computing $\MI$ is that computing $p(\ob_p)$ requires full knowledge of the dynamics of the Markov Game, while computing $p(\ac_p)$ requires marginalizing over all observations (which can be intractable when, e.g., the observation space is high-dimensional). 
Hence, we use a Monte Carlo estimate $\MCMIgen{\policy_p}(\ac_{p,t}; \ob_{p,t})$ of $\MI(\ac_p; \ob_p)$, using data $\brckcur{\brck{\ac_p, \ob_p}: \ac_p \sim \policy_p}$ collected while executing policy $\policy_p$.
Note an unbiased estimator has $\MI = \mathbb{E}_{\bm{\policy}}\brcksq{\MCMI}$.
The utility for a rationally inattentive actor is then computed as $\RI{\utility}_{\timet; i} = \utility_{\timet; i} - \mimult \MCMIgen{\policy_i}(a_{\timet; i}; o_{\timet; i})$.
In practice, even using a single sample to compute $\MCMIgen{}$ is effective, and in particular, estimating $\MI$ using data sampled from the RL policy works well.

\paragraph{Estimating Mutual Information using Discriminators.} 
Given a pair $(a, o)$ for a single agent (omitting subscripts for clarity), we estimate $\MCMIgen{\policy}(a; o)$ from the ratio between $\log p(\ac, \ob)$ (the log-odds under the joint distribution) and $\log p(\ac)p(\ob)$ (the log-odds under the factorized distributions).
This ratio can be estimated using a \emph{discriminator model} $\discriminator_{\policy}(\ac, \ob)$ that learns to classify whether the sample $(\ac, \ob)$ came from the joint or factorized distribution.
We generate samples from $p(\ac, \ob)$ by sampling trajectories using $\policy$, while samples from $p(\ac)p(\ob)$ can be generated by creating random pairs $(\ac, \ob)$ from a batch of sampled trajectories.
This approach is inspired by the structure of Generative Adversarial Networks \cite{goodfellow2014generative}, which use discriminators to learn to synthesize images that are almost indistinguishable from real images.

Our framework is compatible with any MI estimator. 
Our approach is simple and effective, although other MI estimation techniques have also been studied, e.g., \cite{belghazi2018mine}.

As an important point of clarification, this technique for MI estimation can be applied for any pair of variables within some generative process.
We leverage this generality throughout \ourframework{}.

\subsection{Recurrent Policies for Sequential Settings and Heterogeneous Attention Costs} 
\label{sec:policy-architecture}
In addition, we aim to apply our framework in the sequential setting and with heterogeneous attention costs. 
To do so, we use a recurrent model that can summarize historical observations. 
In particular, such models can learn to strategically allocate attention over time.

Before describing details, we provide the following rough sketch of our policy architecture:
\begin{itemize}
    \item The observation is first separated into distinct ``channels''.
    \item For each channel, there is an encoding module which converts its portion of the observation into a noisy encoding.
    \item The encodings from each channel are combined and then fed into a recurrent neural network model to update its hidden state.
    \item The combined encodings and the updated hidden state are fed into a stochastic action module to produce a probability distribution over actions and sample accordingly.
\end{itemize}

\subsubsection{Encoder Modules} 
For each channel, we use the reparameterization trick \cite{kingma2013auto} to compute the mean and standard-deviation of a Gaussian distribution over $y^\chn_t$, which is the noisy encoding of $\ob^\chn_t$:
\eq{
    \mu^\chn_t, \sigma^\chn_t &= \encoder^\chn( \ob^\chn_t, \lstmstate_t ),\\
    y^\chn_t &= \ob^\chn_t + \mu^\chn_t + \sigma^\chn_t \cdot \epsilon^\chn_t,
    \quad \epsilon^\chn_t \sim \mathcal{N}(\bm{0}, \bm{1}).
}
Here $\lstmstate_t$ is the hidden state of the recurrent model (see below), 
and $\epsilon^\chn_t$ is a random sample from a spherical Gaussian with dimensionality equal to that of $y^\chn$ and $\ob^\chn$.

This decomposition allows us to separate out attention costs for distinct channels of information.
In other words, for each channel $\chn$ we want to add an attention cost proportional to the amount of information about $\ob^\chn$ encoded in $y^\chn$.
To that end, analogous to Section \ref{sec:mutual-info-estimation-single-obs}, for each channel $\chn$ we also train a discriminator $\discriminator_\encoder^{\chn}(y^\chn_t, \left[ \ob^\chn_t, \lstmstate_t \right])$ used to estimate $\MCMIgen{m}(y^\chn_t; \left[ \ob^\chn_t, \lstmstate_t \right])$, where brackets denote concatenation.

\subsubsection{Recurrent Stochastic Action Module}
The full encoding $y_t = \left[ y^1_t, \ldots, y^\Chn_t \right]$ of $\ob_t$ concatenates all $\Chn$ encoder samples.
We introduce recurrence by layering a recurrent neural network model that takes $y_t$ and $\lstmstate_t$ as inputs and outputs an updated hidden state $\lstmstate_{t+1}$.
In our experiments, we use LSTMs~\citep{Hochreiter1997}, although many other variations exist in the machine learning literature.
%
%
The encoding $y_t$ and recurrent state $\lstmstate_{t+1}$ are used as inputs to the decoder $\decoder(a_t | y_t, \lstmstate_{t+1})$ which outputs a distribution over actions, from which $\ac_t$ is sampled. 
As with the encoders, we can train a discriminator $\discriminator_{\decoder}(a_t, \left[ y_t, \lstmstate_{t+1} \right])$ used to estimate $\MCMIgen{\decoder}(a_t, \left[ y_t, \lstmstate_{t+1} \right])$, introducing an additional attention cost capturing how much of the encoded information is used to select an action.

\subsection{Training the Principal with Reinforcement Learning}
Given the model structure discussed previously, we now solve the Principal's problem (Equation \ref{eq:pa-problem-principal}) using reinforcement learning (RL) \cite{sutton_reinforcement_2018}.
A core class of RL algorithms train agents by a form of trial-and-error, i.e., by repeated executing a policy $\policy$ in the environment (e.g., the Principal-Agent dynamics) and updating it given the received feedback.
RL has seen many successes in optimizing policies in highly complex settings, such as superhuman performance in Go \cite{silver2017mastering} and Starcraft \cite{alphastarblog}.
Denoting the weights of a policy by $\theta$, we update $\policy{}\brck{\ac | \ob; \theta}$ using policy gradients~\citep{williams1992simple}:
\eq{
    &\Delta \theta \propto \mathbb{E}_{\bm{\policy}, \trans}\brcksq{ 
    \sum_t \nabla_\theta \log \policy ( \ac_t, y_t | \ob_t, \lstmstate_t, \lstmstate_{t+1}; \theta) \RI{\Rew_t}
    },
    \quad
    \RI{\Rew_t} = \sum_{k=0}^{\horizon - t}\gamma^k \RI{\utility_{t+k}},\\ 
    &\log \pi( \ac_t, y_t | \ob_t, \lstmstate_t, \lstmstate_{t+1} ) = \log \decoder(\ac_t | y_t, \lstmstate_{t+1}) + \sum_{\chn=1}^\Chn \log \encoder^\chn (y^\chn_t | \ob^\chn_t, \lstmstate_t), \\
    &\RI{\utility}_t = \utility(s_t, \bm{a}_t) -
    \mimult_\decoder \MCMIgen{\omega}(a_t; \left[ y_t, \lstmstate_{t+1}\right ] ) - 
    \sum_{\chn=1}^\Chn
    \mimult_\chn \MCMIgen{m}(y^\chn_t; \left[ \ob^\chn_t, \lstmstate_t\right ] ).
}
For further details on training and model implementation, see Appendix \ref{section:training-details-appendix} and ~\ref{section:implementation-details-appendix}.
\section{A Principal with a Single Agent in a 1-Step Setting}
\label{Section:Experiments}

\label{section:bandit-experiments}

%
%
In this section, we use \ourframework{} to analyze PA problems with a rationally inattentive Principal.
In particular, we analyze how the degree of the Principal's inattention changes outcomes.

First, we confirm that \ourframework{} is sound and matches theoretical predictions of optimal payment structures by \cite{mirrlees1976optimal}.
Our results show how Principal inattention leads to simpler but less profitable payment structures.
Furthermore, using \ourframework{}, we can give more quantitative descriptions of the relationships between the parameters of the PA model and overall outcomes.

Second, we show and quantify several non-trivial consequences of rational inattention:
\begin{enumerate}
    \item that the effects of the Principal's inattention depend on the type of the Agent,
    \item that if Agents are prone to sub-optimal effort choices, payment schedules are more sensitive to the Principal's attention cost, and
    \item that Agent (but not Principal) welfare can improve if the Principal is inattentive.
\end{enumerate}

\subsection{Optimal Payments: Definitions and Problem Statement}

We first evaluate in a PA setting with a single Principal and a single Agent, following classic work \citep{mirrlees1976optimal, Holmstrom1987, Spremann1987, haubrich1994risk}.
Here, a manager (the Principal) must choose how much to pay its employee (the Agent), contingent on the output of its labor.
The Agent can influence this output by choosing how much effort to use.
An information asymmetry exists: the Principal cannot directly observe the Agent's effort choice and must instead base its payment contract on the output.
%
%

In our experiments, the Principal also pays a cost to observe information about the output.
This scenario is inspired by \citet{mirrlees1976optimal}, which analyzed optimal payment schedules when the Principal also controls the noise level added to the (perceived) output it uses to set payments.
Here, the Principal acts first by setting the parameters of the payment schedule $\schedule$.
Second, the Agent chooses how much effort $\effort$ to use based on the chosen schedule.
Finally, the Principal and Agent utilities are determined based on the resulting output and payment.

\paragraph{Labor-Effort Model.}
The Agent's labor output $\output$ is a stochastic function of its effort action $\effort$.
Specifically:
\begin{equation}
    p(\output|\effort) = 
\begin{cases}
  0.7 & \text{if } \output = \effort, \\
  \frac{0.7^{|\output-\effort|}}{0.3 \cdot c} & \text{otherwise},
\end{cases}
\end{equation}
where $c$ is a normalization factor such that $p(\output|\effort)$ is a proper probability distribution satisfying $\sum_{\output \in \Output} p(\output|\effort) = 1$. 
Under this Agent policy, effort $\effort$ and output $\output$ are highly correlated, with the probability of output $\output$ decreasing exponentially as the difference between $\effort$ and $\output$ grows.

\paragraph{Employee Payment Model.}
The Principal sets a payment schedule $\schedule$, which yields payment $\payment$ as a stochastic function of output $\output$, i.e., $\payment \sim \schedule(\output) = \mathcal{N}(\mu_\output, \sigma_\output)$.
The Principal controls the pay schedule by choosing, for each possible output level $\output$, the mean $\mu_\output$ and standard deviation $\sigma_\output$ of the stochastic payment.
While stochasticity is perhaps an unnatural feature of typical payment contracts, here the stochasticity aims to model the Principal's uncertainty over $\output$, e.g., due to limited attention.
By assumption, the Agent accurately anticipates the shape of the payment schedule $\schedule$ when making its decisions, e.g., based on prior experience or reputation.

\paragraph{Principal and Agent Objectives.}
The utility of the Principal ($\utility_p$) and Agent ($\utility_a$) follow standard economic utility functions:
\begin{equation}
    \utility_a(\payment, \effort; \rho) = 
    \underbrace{
    \texttt{CRRA}(\payment; \rho)
    }_{\text{Income Utility}} - 
    \underbrace{
    \effort
    }_{\text{Work Disutility}},
    \quad
    \underbrace{\utility_p(\payment, \output)}_{\text{Profit}} = 
    \underbrace{
    \output
    }_{\text{Revenue}} - 
    \underbrace{
    \payment
    }_{\text{Amount Paid}}.
\end{equation}
Here \texttt{CRRA} is the concave Constant Relative Risk Adverse function~\citep{pratt1978risk}; the risk aversion parameter $\rho$ sets its concavity (we use $\rho = 2$).

\paragraph{Agent Policy.}
In this 1-step PA setting, the flexibility of our \ourframework{} framework enables us to analyze how the effects of Principal inattention depend on the characteristics of the Agent.
To do so, we model the Agent's policy $\policy_a$ as a logistic quantal response function~\citep{McKelvey1995qre}:
\eq{\policy_a(\effort | \schedule; \beta) = \frac{e^{\beta \utility_a(\effort | \schedule)} }{\sum_{e'}e^{\beta \utility_a(\effort' | \schedule)}},
}
where $\utility_a(\effort | \schedule)$ denotes the \emph{expected} utility of effort $\effort$ given schedule $\schedule$.
When $\beta=\infty$, this models an Agent that perfectly selects the (utility-maximizing) best response $\effort$ given $\schedule$.
When $\beta$ is finite, this embeds the common intuition that the Agent will not \textit{always} select the utility-maximizing effort, but will tend to choose more rewarding levels of effort.
The parameter $\beta \geq 0$ controls the strength of this tendency, with $\beta=\infty$ yielding deterministic, utility-maximizing effort and $\beta=0$ yielding random effort choices.
In our experiments below, we examine the effects of Principal inattention under different choices of Agent $\beta$.

Note that \ourframework{} does not \textit{prescribe} this choice of Agent policy model. 
For example, in Section~\ref{section:sequential-experiments}, the Agent policy is a general recurrent neural network.
However, in the single-step setting, training $\policy_a$ using RL is not necessary since $\utility_a(\effort | \schedule)$ and the optimal $\policy_a$ can be calculated directly once $\schedule$ is known.

\paragraph{Solving the Principal's Problem.}
We apply \ourframework{} to learn the optimal payment schedule:
\eq{\label{eq:planner-problem}
    \schedule^* = \argmax_\schedule \mathbb{E}_{
    \payment \sim \schedule(\output),
    \output \sim p(\output | \effort),
    \effort \sim \policy_a(\effort | \schedule; \beta)
    } \brcksq{ \utility_p(\payment, \output) - \mimult \MCMI_{\schedule}(\payment; \output)}.
}
%
Here, $\mathbb{E} \left[ \utility_p(\payment, \output) - \mimult \MCMI_{\schedule}(\payment; \output) \right]$ describes the objective of the Principal;
$\mathbb{E} \left[ \utility_p(\payment, \output) \right]$ is the expected profit and $\mathbb{E} \left[ \mimult \MCMI_{\schedule}(\payment; \output) \right] = \mimult \MI_{\schedule}(\payment; \output)$ is the attention cost for $\schedule$.
%
%
%

We use $N$-sample rewards to compute policy gradient estimators to learn the distribution parameters $\mu_{\output}, \sigma_{\output}$ of $\schedule^*(\output)$:
\eq{
& \RI{\rew}_\schedule = \frac{1}{N} \sum_{n=1}^N \brcksq{ \utility_p(\payment_n, \output_n) - \mimult \MCMIgen{\schedule}(\payment_n; \output_n)
}, \\
& \payment_n \sim \schedule(\output_n), \quad
\output_n \sim p(\output | \effort_n), \quad
\effort_n \sim \policy_a(\effort | \schedule; \beta).
}
We compute the \ourframework{} reward for a given schedule $\schedule$ as follows.
First, we compute $\utility_a(\effort | \schedule)$ for each $\effort$ to get $\policy_a(.|\schedule; \beta)$.
We then take $N$ samples of $\effort$ from $\policy_a$.
For each sample $\effort_n$ we sample an output $\output_n \sim p(\output | \effort_n)$, and finally a payment $\payment_n \sim \schedule(\output_n)$.
The reward is the average of 
$\utility_p(\payment_n, \output_n) - \mimult \MCMIgen{\schedule}(\payment_n; \output_n)$ across the $N$ pairs of $(\payment_n, \output_n)$.

Recall that the Principal's attention cost $\mimult \MI_{\schedule}(\payment; \output)$ is proportional to the amount of information that the payment $\payment$ provides about the output $\output$.
An intuitive interpretation is that this cost captures how carefully the Principal would need to attend to the actual output when paying the Agent.

For example, suppose that under a schedule $\schedule$ the likelihood of the Agent receiving a given payment $\payment$ does not change much with the Agent's output $\output$.
We can interpret this as the Principal not paying close attention to the Agent's output when determining its payment, which yields a small $\MI_{\schedule}(\payment; \output)$.

\subsection{Results}
\subsubsection{Comparison With Prior Literature.}
When analyzing a similar setting to the one considered here, \citet{mirrlees1976optimal} found the theoretically optimal pay schedule should be of the form 
\eq{
\schedule(\output) = \max(A\output + B, C)^\frac{1}{\rho}.
}
Despite some differences in our modeling choices,\footnote{
\citet{mirrlees1976optimal} models costly Principal inattention by allowing the Principal to choose $\theta$, which controls the magnitude of the noise in the observed output signal, and where $\theta^2$ is subtracted from the utility as an attention cost.
While we model attention using mutual information, both models regard inattention as an endogenous source of noise.
One important difference, however, is that our work considers Agent behavior which incorporates a notion of bounded rationality.
}
our framework discovers optimal pay schedules that match closely with this work:
when fitting $A$, $B$, $C$ and $\rho$ to our data, we measure $r^2 \geq 0.98$ for all $\beta$.
The best fit $\rho$ was close to the theoretical value for some values of $\beta$ (true $\rho$: 2, best fit $\rho$: 2.19 for $\beta = 5$). 
Interestingly, we do observe that the best fitting $\rho$ tends to increase with $\beta$ in our model.

Notably, the influence of Agent sub-optimality (encoded in $\beta$) was outside the scope of this prior work.
Here, we can in fact observe this relationship between $\rho$ and $\beta$ using \ourframework{}.
Furthermore, our approach enables novel insight into how optimal payment schedules depend on Principal inattention and boundedly rational Agent behavior.
We explore these insights below by analyzing the optimal payment schedules discovered by \ourframework{}.

\subsubsection{Wage Schedules under Inattention.}
\begin{figure}[t]
     \centering
     \includegraphics[width=\textwidth]{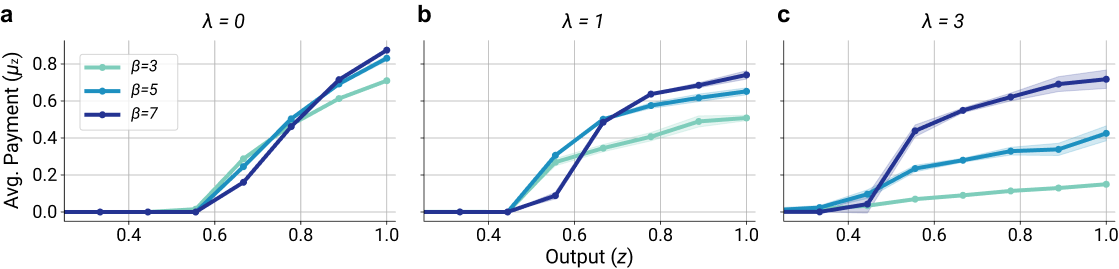}
\caption{ 
\textbf{A comparison of pay schedule means ($\mu_\output$'s) across different Agent policy $\beta$ values (indicated by color).}
Higher $\beta$ increases the likelihood that the Agent selects its utility-maximizing effort.
\textbf{(a)} Schedules when there is no attention cost ($\mimult = 0$).
\textbf{(b, c)} Same as (a), but with attention costs of $\mimult=1$ and $\mimult=3$, respectively.
All results were averaged across 5 random seeds; shading denotes 95\% confidence regions.
}
\label{fig:ps_temp}
\end{figure}
%

Figure~\ref{fig:ps_temp} compares payment schedules (specifically, mean payment $\mu_\output$ for each output $\output$) across different Agent policy $\beta$ values.
In general, larger attention costs encourage flatter schedules.
This tendency corroborates the intuition that attention costs encourage \textit{simpler} policies, where policy outputs (in this case, payments) depend less precisely on policy inputs (in this case, output).
We interpret this simplicity as satisfying the limitations on how much information the Principal can process, i.e. its bounded rationality.

Interestingly, when the Principal does not face any attention cost (Fig.~\ref{fig:ps_temp}a), optimal pay schedules only weakly depend on the choice characteristics of the Agent, i.e. its tendency to towards random (low $\beta$) versus optimal (high $\beta$) effort choices.
However, as attention becomes more costly, differences between schedules under different Agent $\beta$ values become more pronounced (Fig.~\ref{fig:ps_temp}c).
This trend is primarily driven by the way Principal attention costs influence the optimal schedule when the Agent is more random.
That is, lower Agent $\beta$ leads to optimal schedules that ``flatten'' more quickly in response to increasing Principal attention cost.


\subsubsection{Effects of Agent Behavior.}
We provide the following explanation for the observed payment schedules.
The Agent's $\beta$ parameter determines how sensitive its choice of effort is to the differences in their expected utilities.
When the Agent's choices are relatively \textit{insensitive} (e.g., $\beta = 3$), influencing the Agent's behavior via $\schedule$ means paying more attention.
In other words, the less optimal the Agent's behavior, the weaker the Principal's incentive to pay attention.

\begin{figure}[t]
    \centering
    \includegraphics[width=0.95\textwidth]{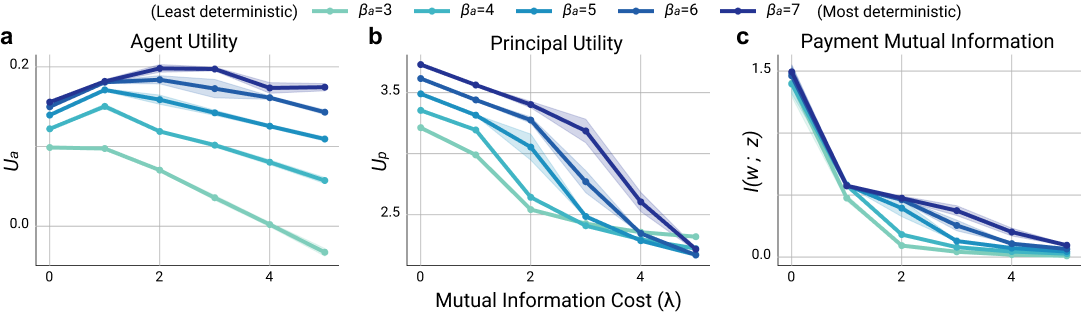}
    \caption{
    \textbf{Interaction effects between the Principal's attention cost ($\mimult$, x-axis) and the Agent policy $\beta$ values (indicated by color).}
    \textbf{(a, b)} Agent and Principal utility.
    \textbf{(c)} Mutual Information between output $\output$ and payment $\payment$.
    All results were averaged across 5 random seeds; shading denotes 95\% confidence regions.
    }
    \label{fig:diff_temps}
\end{figure}

An examination of Principal utility (i.e. profit) and $\MI(\payment; \output)$ (i.e., attention) corroborates this insight.
Figure~\ref{fig:diff_temps} show how these quantities change with increasing attention costs, separately for each Agent $\beta$ value.
As expected, attention costs limit the profit the Principal is able to extract, with costlier attention leading the Principal to prefer simpler but less profitable payment schedules.
 
Interestingly, profits fall off more quickly with sub-optimal Agent behavior.
This is reflected in the amount of attention spent by the Principal as attention costs increase.
In general, the Principal chooses to spend more attention when the Agent is more optimal (higher $\beta$).

%

\subsubsection{Agent Utility under Inattention.}
%
%
%
Surprisingly, we observe that the Agent's utility $\utility_a$ can actually increase when the Principal is inattentive, and that higher $\beta$ tends to increase the beneficial range of inattention (Fig.~\ref{fig:diff_temps}a).
%
That the Agent can benefit at all from inattention is counter-intuitive when considering that the optimal 0-attention schedule pays 0 for all outputs.
Nevertheless, our framework reveals that Principal inattention can improve Agent utility (which is not true for the Principal's own utility).
To our knowledge, this is a novel finding.

\subsubsection{Comparing MI and Entropy Costs.}
\begin{figure}[t]
    \centering
    \includegraphics[width=0.95\textwidth]{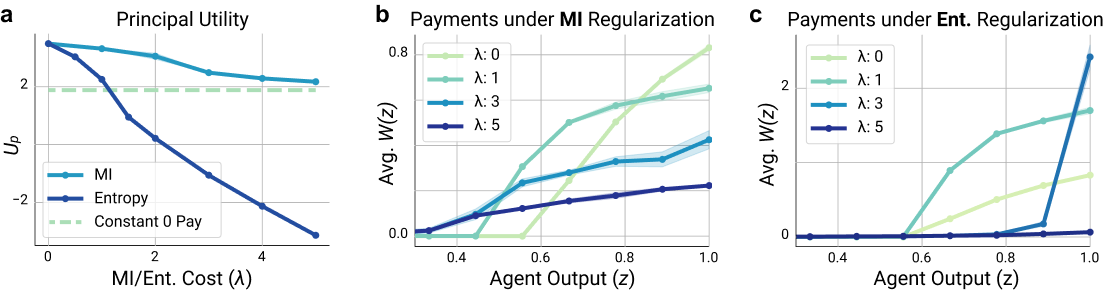}
    \caption{
    \textbf{Comparing MI-based versus Entropy-based regularization.}
    \textbf{(a)} Principal utility as a function of cost/regularization strength $\mimult$. The constant 0-pay schedule provides a meaningful lower bound.
    \textbf{(b, c)} Pay schedules ($\mu_z$ for each output $\output$) under MI (b) and Entropy (c) regularization.
    All results are averaged over 5 random seeds and plotted with 95\% confidence intervals. 
    }
    \label{fig:bandit_res}
\end{figure}
As an additional validation of our technical framework, we present a comparison of outcomes when using attention costs based on our \ourframework{} framework (i.e. using MI) versus when using entropy regularization.
%
%
%
%
%
MI-based and entropy-based regularization are closely related~\citep{grau2018soft, leibfried2020mutual}, with subtle but important differences%
\footnote{
    Under some conditions, they are identical. 
    However, the key difference is that entropy regularization penalizes the policy for deviating from a \emph{fixed, uniform} prior, whereas RI-style MI regularization penalizes deviations from an \emph{optimal} prior.
}.
Nevertheless, they yield markedly different outcomes (Figure~\ref{fig:bandit_res}).

To highlight an important difference, consider the constant 0-pay schedule.
This provides no incentive for the Agent to spend effort and costs the Principal nothing, and it therefore provides a reasonable lower bound on $\utility_p$.
This lower bound should hold under bounded rationality, since no attention is required when $\schedule$ treats all outputs identically.
Indeed, under \ourframework{}, $\utility_p$ approaches this lower bound as increasing the strength of the attention cost $\mimult$ leads the Principal to trade more profitable $\schedule$'s for ones with smaller demands on attention (Figs.~\ref{fig:bandit_res}a, b,~\ref{fig:diff_temps}c).
In contrast, under entropy regularization, increasing $\mimult$ quickly yields $\schedule$'s that violate the $\utility_p$ lower bound (Fig.~\ref{fig:bandit_res}a, c).

\section{Sequential Setting with Multiple Agents and Information Channels}
\label{section:sequential-experiments}
We now extend the analysis to a more general and complex \emph{sequential} setting featuring multiple Agents (e.g., a team of employees) and multiple channels of information.
We show that \ourframework{} enables a detailed analysis of the relationships between information channels and their effects on inattention, and this (disparately) impacts Agents of different types.

In this scenario, a rationally inattentive Principal sets wages for each Agent and has access to $\Chn$ heterogeneous information channels. 
For example, a Principal may pay a cost to observe individual outputs, which may correlate with an Agent's hidden ability level.
This setting combines several themes examined in prior work, such as teams~\citep{holmstrom1982moral}, signalling~\cite{Spence1973}, and learning through repeated interaction~\citep{holmstrom1999managerial,Alos-Ferrer2012}. 
By extending these concepts, \ourframework{} yields new insights on the implications of the Principal's inattention and reveals a rich spectrum of effects:
\begin{enumerate}
    \item inattention can have opposing consequences depending on the channel of information,
    \item inattention to Agents' outputs closes wage gaps based on ability differences,
    \item conversely, inattention to Agents' efforts induces a social dilemma dynamic in which Agents work harder, but essentially for free, and 
    \item through both effects, inattention mediates a trade-off between the utility of the Principal and the average utility of the Agents.
\end{enumerate}

\subsection{Definitions and Problem Setting}
Our environment features one Principal and a team of $\numa=4$ Agents.
We assume there are $\Type=5$ possible Agent abilities $\ability$.
Each Agent's ability is sampled randomly at the start of each episode and the Principal cannot directly observe Agents' abilities.
Each episode proceeds for a duration of $\horizon$ timesteps.
On each timestep $t$, the Principal acts first and sets a wage $\payment_{i,t}$ for each Agent $i$.
Each Agent moves second: it knows $\wage_{i,t}$ before choosing hours $\hours_{i,t}$ and effort $\effort_{i,t}$, and earning income $\payment_{i,t}\cdot\hours_{i,t}$.
These two choices (hours $\hours_{i,t}$ and effort $\effort_{i,t}$), along with the Agent's ability $\ability_i$, determine its output $\output_{i,t}$ for that timestep:
\eq{
\output_{i,t} = \hours_{i,t} \cdot (\ability_i + \effort_{i,t}).
}
We emphasize (1) that, for a given amount of hours and effort, output increases with ability, and (2) that effort is a substitute for ability.

%

\paragraph{Objectives.} 
As before, Principal utility $\utility_p$ measures profit:
\eq{
\underbrace{\utility_p(\bm{\wage}, \bm{\hours}, \bm{\output})}_{\text{Profit}} = 
    \underbrace{
    \sum_{i \in [n_a]} \output_i
    }_{\text{Revenue}} - 
    \underbrace{
    \sum_{i \in [n_a]}\wage_i \hours_i
    }_{\text{Wages Paid}}.
}
We define Agent utility $\utility_a$ following standard utility functions, where the optimal choice of hours $\hours$ increases with $\wage$.
The Agent utility is:
%
\begin{equation}
    \utility_a(\wage, \hours, \effort) = 
    \underbrace{
    \texttt{CRRA}(\wage \cdot \hours; \rho)
    }_{\text{Income Utility}} - 
    \underbrace{
    c_{\labor} \hours^\alpha (1+\effort)
    }_{\text{Work Disutility}},
    %
    %
\end{equation}
where $\rho, c_{\labor}$ and $\alpha$ are constants governing the shape of $\utility_a$.

Our results will show that these definitions imply that the profit-maximizing wage $\wage_i$ for Agent $i$ increases with its ability $\ability_i$.
As such, the Principal benefits from accurately inferring an Agent's ability and each Agent benefits from receiving a higher wage.
While ability is hidden, the Principal can \textit{infer} it over the course of multiple interactions, but this inference is sensitive to what the Principal pays attention to.
Consequently, inattention can create endogenous information asymmetries through which strategic Agents can manipulate the Principal's inference and hence their own wages.

%
For instance, note that, while effort $\effort_i$ increases an Agent's work disutility without increasing its income, Agents may still choose to exert non-zero effort $\effort_i > 0$.
To provide an intuition for this, recall that spending effort increases output rate ($\output_i$ per hour $\hours_i$), resembling higher ability $\ability_i$.
%
%
The Principal can't see each $\ability_i$ but can see the output $\output_i$ and hours $\hours_i$.
Hence, output rate is a ``signal''~\citep{Spence1973} of ability and Agents can misrepresent their ability by spending effort.
As such, a strategic Agent may choose to spend effort on timestep $t$ in order to enjoy a higher wage on timestep $t+1$.
By extension, any such incentive to spend effort goes away by the end of the episode.
These aspects of the PA setting inform the effects of effort inattention, which we examine below.

\paragraph{Attention Costs}
Because a strategic Principal must \textit{infer} private Agent features (e.g., ability), outcomes will depend on any inference costs the Principal experiences (e.g., attention costs).
To capture these dependencies, we model the Principal policy using the \ourframework{} architecture described in Section~\ref{sec:policy-architecture}.
%
We also wish to model Agents that behave strategically; as such, we use a similar, recurrent policy architecture to model the Agents.\footnote{
However, to isolate and explore the effects of distinct \textit{Principal} attention costs, we do not impose any attention costs on the Agents.
Therefore, the Agents' reward during training is only their utility $\utility_a(\wage_{i,t}, \hours_{i,t}, \effort_{i,t})$.
}

\begin{figure}[t]
    \centering
    \includegraphics[width=0.95\textwidth]{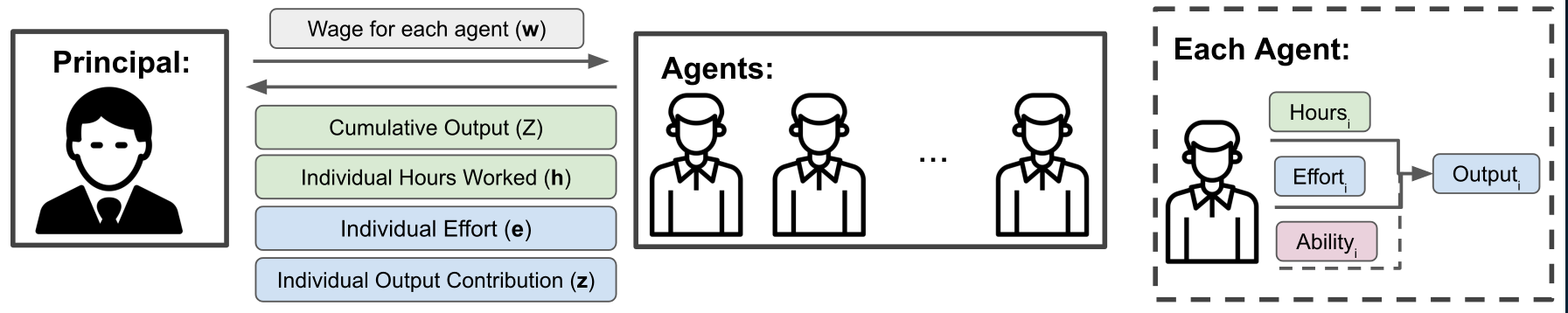}
    \caption{
    \textbf{Visualizing a single timestep in an episode of our Sequential Multi-Agent Setting.}
    Green variables (such as cumulative output) are not costly while blue variables (such as individual output) are costly for the principal to observe. 
    The red variable (Ability) cannot be observed.}
    \label{fig:pa_env}
\end{figure}

For the Principal, we model $\Chn=3$ information channels: one is ``easy'' and low-cost to observe and two are ``hard'' and high-cost. 
The low-cost channel $\ob^f_p$ includes information that we regard as freely available $(\mimult^f = 0)$, e.g., the time $t$, the hours worked $\bm{\hours}$ (workers often fill out timesheets which makes $\hours_i$ easy to see), and the \emph{total} output, $Z = \sum_{i \in [\numa]}\output_i$ (managers can see the final result).
However, it is high-cost to see \emph{individual} contributions: we use a high-cost channel $\ob_p^\effort$ for efforts $\bm{\effort}$, and a high-cost channel $\ob_p^\output$ for outputs $\bm{\output}$.
%
This models a Principal who can spend time and attention to observe individual Agents to reduce uncertainty about their true ability, e.g., their working styles and productivity (Figure~\ref{fig:pa_env}).
Modeling the attention cost of output and effort separately lets us study the effects and interactions of unequal observation costs.
The Principal's reward is
\begin{equation}
    \RI{\utility}_{p,t} = \utility_p(\bm{\wage}_t, \bm{\hours}_t, \bm{\output}_t) - 
    \underbrace{\mimult^\output \MCMI(y^\output_t; \bm{\output}_t)
    }_{\text{Individual Output and}} - \underbrace{\mimult^\effort \MCMI(y^\effort_t;\bm{\effort}_t)
    }_{\text{Effort Attention Cost}}.
\end{equation}
Here, $y_t^\output$ and $y_t^\effort$ come from the Principal's \ourframework{} policy architecture: they are the (stochastic) encodings of outputs $\bm{\output}_t$ and efforts $\bm{\effort}_t$, respectively, and the Principal policy selects wages $\bm{\payment}_t$ on the basis of these encodings.
Thus, the Principal's bounded rationality is modeled through the cost to get information about effort and individual outputs.
The Principal's \ourpolicy{} architecture would also let us use attention costs $\MCMI(y^f_t;\ob^f_t)$ and $\MCMI(\bm{\wage}_t; y_t)$, but we omit those here%
\footnote{
We do add entropy regularization over $\omega(a|y)$ for both the Principal and Agent to encourage exploration.
}.
%
%
%
%
Additional training details are in the appendix.

\subsection{Results}
We highlight several noteworthy observations:
(1) The temporal patterns of Principal attention reflect the value of information over time;
(2) Principal inattention to individual outputs and inattention to efforts have distinct, nearly opposite, consequences on actor utilities;
(3) Principal inattention to effort drives Agents to spend more effort, resulting in their lower overall utility.

\subsubsection{The \ourframework{} Principal Learns the Value of Information}
\begin{figure}[t]
     \centering
     \includegraphics[width=0.95\textwidth]{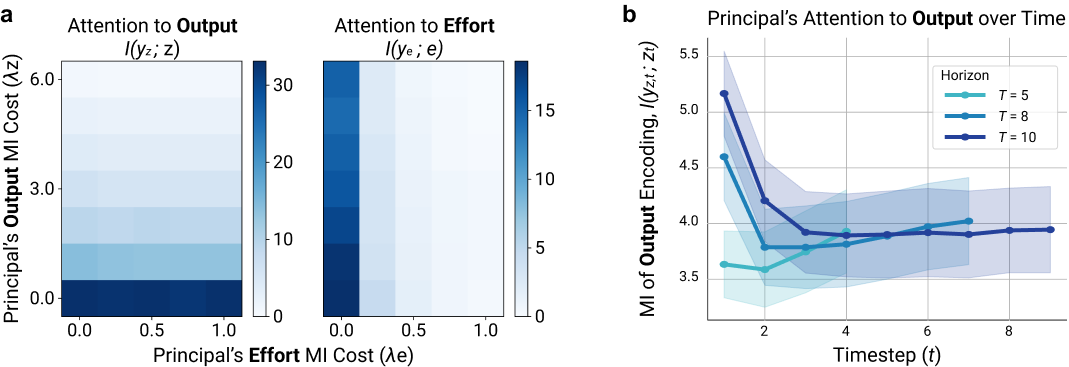}
    
\caption{
\textbf{The Principal's attention patterns.}
\textbf{(a)} Heatmaps illustrate the amount of attention paid by the Principal to output $\output$ (left) and effort $\effort$ (right), as measured in the MI between the given observation channel and its encoding. Higher MI costs lead the Principal to pay less attention to the associated observation.
\textbf{(b)} Attention paid by the Principal to output $\output_t$ at each timestep (measured as the MI between the actual output $\output_t$ and encoded output $y^\output_t$), for 3 different episode lengths. Information about output is most valuable at the start of the episode, with that value increasing for longer episodes.
All results are averaged over 20 random seeds; shading denotes 95\% confidence intervals.
%
}
\label{fig:ma_attention}
\end{figure}
As a starting point for our analysis, we confirm that the Principal's \ourframework{} policy architecture learns to allocate attention in keeping with basic intuitions about its value.
As shown in Figure~\ref{fig:ma_attention}a, the attention paid by the Principal to output depends on the associated attention cost, similarly for effort.
We also point out a tendency for the Principal to spend less attention on effort when the cost of attending to output is high.
This makes intuitive sense, since effort provides less information about ability when the output is uncertain.

Figure~\ref{fig:ma_attention}b shows how much attention the Principal pays to output over time, i.e., $\MCMI(y^\output_t; \bm{\output}_t)$ throughout the episode.
Attention is highest at the beginning of the episode and then decreases over time. 
Intuitively, it is most efficient for the Principal to pay attention to outputs at the beginning of the episode, as this information can be used throughout the episode (Agent abilities do not change).
Additionally, the initial spike is larger for longer horizons, as initial information is more valuable the longer it can be reused.

\subsubsection{Welfare Implications of Principal Inattention.}
\begin{figure}[t]
    \centering
    \includegraphics[width=0.75\textwidth]{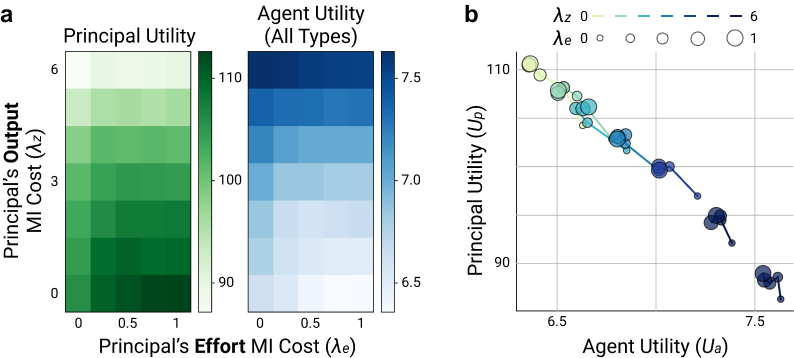}
    \caption{
    \textbf{Principal and Agent utilities under inattention.}
    \textbf{(a)} Heatmaps illustrate the Principal utility (left) and average Agent utility (right) and for each $\olamb$ and $\elamb$.
    \textbf{(b)} A scatterplot representation of the Principal/Agent utilities in (a). Color denotes the cost of output attention and dot size denotes the cost of effort attention.
    Inattention to output decreases Principal utility while increasing (average) Agent utility, with the opposite being true for inattention to effort.
    All results are averaged over 20 runs and generated for episodes with horizon $\horizon = 5$. 
    %
    %
    %
    %
    }
    \label{fig:mamt_res1}
\end{figure}
We analyze welfare related to actors' utilities.
Principal and Agent utilities are negatively correlated across varying levels of $\olamb$ and $\elamb$ (Figure~\ref{fig:mamt_res1}).
Note that the model with $\olamb = \elamb = 0$ implies a ``rational'' Principal. 
This indicates that the Principal's inattention (e.g., reflecting bounded rationality) has opposing implications for the Principal and Agents. 
From Figure~\ref{fig:mamt_res1}, we see that Agents' average utility increases and the Principal's utility decreases when the Principal's attention cost for individual outputs $\olamb$ increases.
Conversely, Agent utility decreases with increasing Principal attention cost on effort $\elamb$. 
We discuss the causes for these trends below.

\begin{figure}[t]
    \centering
    \includegraphics[width=0.95\textwidth]{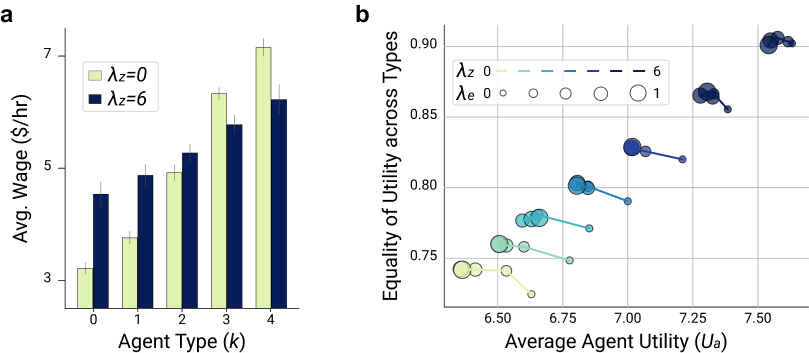}
    \caption{
    \textbf{Wages, equality, and utility trade-offs in the sequential setting.}
    \textbf{(a)} Average wage for each Agent ability type, comparing when the Principal does not face any cost for attending to output ($\olamb = 0$, yellow) versus when this cost is at the highest tested value ($\olamb = 6$, blue). Error bars denote to 95\% confidence intervals.
    %
    \textbf{(b)} Equality of utility across Agent types vs. average Agent utility, for each $\olamb$ and $\elamb$.
    Inattention to output drives wages to be more similar and higher overall, resulting in more equality in the utilities experienced by Agents of different ability types \textit{and} higher average utility overall.
    Inattention to effort tends to reduce Agent utility with only small gains in equality.
    All results are averaged over 20 runs and were generated with $\horizon = 5$.
    %
    }
    \label{fig:mamt_res2}
\end{figure}

The Principal has a different optimal wage for each ability $\ability$ as shown in Figure~\ref{fig:mamt_res2}a. 
When the Principal is fully rational it can use output and effort to accurately infer ability.
Similar to what was seen in Section~\ref{section:bandit-experiments}, an increase in attention costs (in this case, on output $\olamb$) leads the Principal to set wages in a manner that is less profitable but also less attentionally demanding.
Inattention to output results in more uncertainty over each Agent's ability; the Principal compensates with a ``better safe than sorry'' approach where wages are more similar between ability types but higher on average (ostensibly because it is better to over-pay low-ability Agents than under-pay high-ability Agents).
In sum, the same force that creates a lower-utility equilibrium for the Principal has higher average utility for the Agents.

\subsubsection{Implications of Inattention Differ by Agent Ability.}
While the \textit{average} Agent utility increases with $\olamb$, it does not increase for all Agent types.
Specifically, the utility of the (highest-) lowest-ability Agent's (decreases) increases. 
Hence, the Principal's uncertainty over individual outputs decreases the wage (and utility) differences between Agents of different ability.
This is particularly relevant when considering the \textit{equality} of utility.
A common inequality metric is the Gini coefficient~\citep{gini1971variability}, computed as a normalized sum of income differences; equality can be defined as $\texttt{eq} = 1 - \frac{N}{N - 1}\texttt{gini}$.
Figure~\ref{fig:mamt_res2}b shows that Agents' average utility \textit{and} equality across types both increase for higher output attention cost $\olamb$. %

\begin{figure}[t]
    
     \centering
     \includegraphics[width=\textwidth]{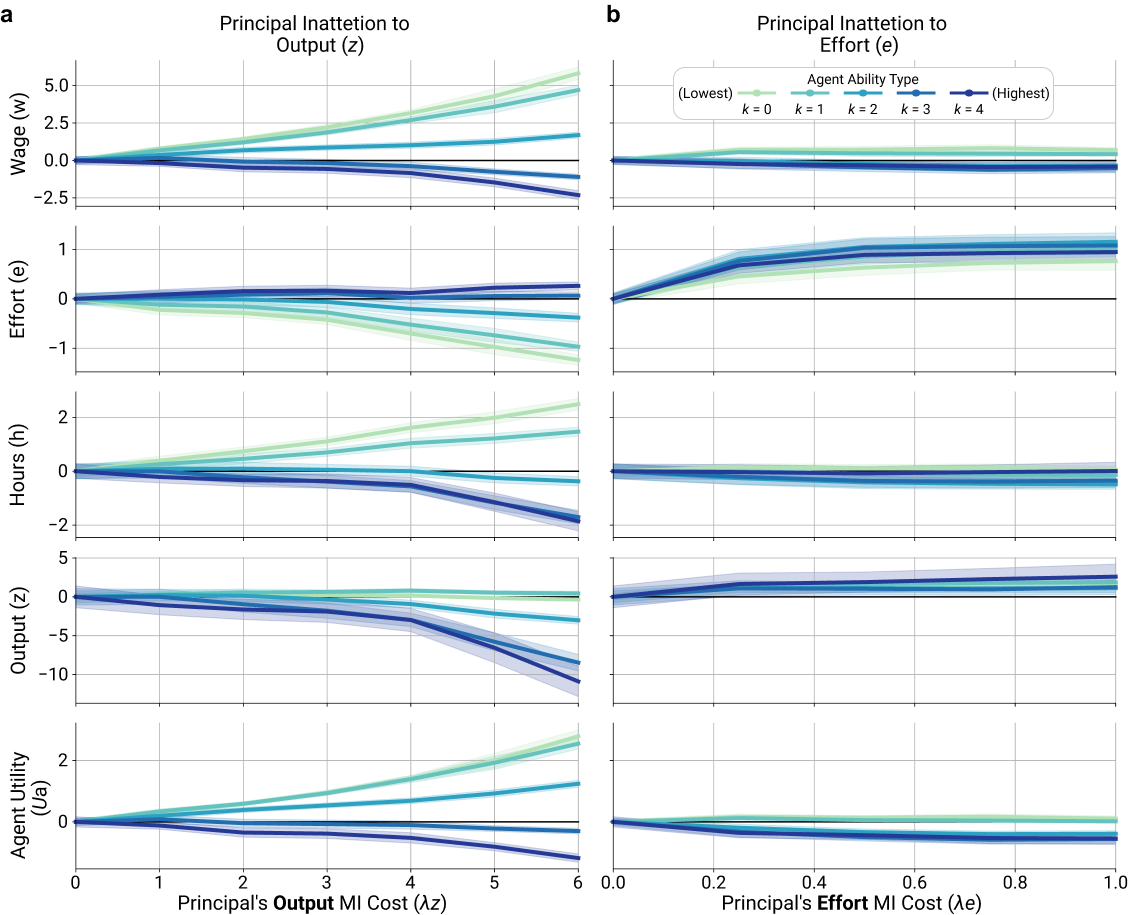}
\caption{
\textbf{Breakdown of inattention effects for each Agent ability type.}
Plots illustrate how wages, effort used, hours worked, output generated, and utility experienced (from top to bottom, respectively) change with increasing inattention to output \textbf{(a)} and effort \textbf{(b)}.
Each colored line represents the change (relative to the $\mimult=0$ condition) experienced by Agents of a specific ability type.
Both (a) and (b) consider the respective inattention type in isolation: the results in (a) use $\elamb=0$ and the results in (b) use $\olamb=0$.
All results are averaged over 20 random seeds, using $\horizon = 5$; shading denotes 95\% confidence intervals. 
%
}
\label{fig:agent_trends}
\end{figure}
One advantage of our framework is that it allows us to examine the consequences of inattention directly, including at the granularity of individual Agent ability types.
Figure~\ref{fig:agent_trends} illustrates how Principal inattention (separately, to output and to effort) drives changes in the wages, effort, hours, and utility for each such Agent type.
We are able to directly see this differential effect that inattention to output has on Agent utility.
Furthermore, diverting wages from higher- to lower-ability Agents results in net losses to output, which explains why this form of inattention leads to less profit for the Principal.

\subsubsection{Inattention to Effort induces Signalling and Social Dilemma Dynamics.}
Recall that effort and ability have the same impact on an Agent's output.
When $\elamb$ increases, it becomes costly for the Principal to distinguish output due to the Agent's ability from output due to its effort.
Interestingly, this leads to \textit{signaling equilibria} in which Agents choose to use more effort (Fig.~\ref{fig:agent_trends}b), resulting in lower average Agent utility and higher Principal utility (Fig.~\ref{fig:mamt_res1}a).
%
%

To build an intuition for this outcome, imagine the hypothetical case where each episode features only two Agents, one with low ability and one with high ability. 
If both Agents spend the same level of effort, the Principal can identify their ability levels on the basis of individual output, and the higher-ability Agent will enjoy a higher wage as such.
If the Principal does not attend to effort, the lower-ability Agent gains an incentive to spend additional effort early on so that its output resembles that of the higher-ability Agent.
Based on our above analysis of how output inattention affects wages (Fig.~\ref{fig:mamt_res2}a), the additional effort spent by the lower-ability Agent would increase its own wages and decrease the wages of the higher-ability Agent (because the Principal cannot distinguish them).
In response, however, the higher-ability Agent can also increase its effort, which makes individual outputs again indicative of ability, and restores the original wages that benefit the higher-ability Agent.

The key here is that, if all Agents increase (or decrease) effort by a similar amount, the Principal's ability to infer ability on the basis of individual output remains the same.
In effect, the dynamic described above predicts that inattention to effort will ensure that each individual Agent faces an incentive to increase its effort, but when all the Agents do so it will have negligible impact on wages.
Put another way, inattention to effort creates a type of \textit{social dilemma} amongst the Agents where collective Agent welfare is maximized by behaviors (i.e. no changes in efforts) that contradict individual incentives.

This explains the observed behavioral shifts induced by inattention to effort (Fig.~\ref{fig:agent_trends}b), where, as attention to effort becomes costlier for the Principal, Agents increase their effort in similar measure but wages remain stable; the result is higher output (which the Principal enjoys essentially for free) and lower Agent utility overall.
Interestingly, the effects of effort being costly to observe are nullified if the individual outputs are also hard to observe (Fig.~\ref{fig:mamt_res2}b), which also agrees with our above interpretation.

Why then doesn't the Principal \textit{always} ignore effort?
This analysis perhaps suggests that it should.
However, \textit{within an individual episode} the Principal can benefit from correctly inferring ability, hence there is an incentive to attend to effort.
It is only through this analysis that we see that the Principal would prefer the social dynamic created by resisting this incentive.


\parwrapper{Limitations.} 
While RI is a general model of boundedly rationality, it does not cover all models of human irrationality. 
For simulations that wish to use more specialized models of bounded rationality, \ourframework{} may not be sufficient. 
Examples include hyperbolic discounting~\citep{kirby1995preference} to model time-inconsistent delay discounting (humans tend to prefer rewards in the near future much more than rewards farther in the future) or prospect theory~\citep{Tversky1992} to model loss-aversion.
Future research may extend \ourframework{} to include such notions of bounded rationality.
\section{Conclusion}

\paragraph{Future Work.} 
The effects of bounded rationality in multi-agent systems are highly relevant in understanding complex economic behaviors in the real world. 
However, modeling and measuring bounded rationality and extracting economic insights, e.g., based on equilibrium assumptions, still pose significant technical and practical challenges.

Towards progress in this domain, our results show that \ourframework{} and reinforcement learning more generally can be useful tools for studying rich behavioral models and their impact on economic interactions.
RL is a general learning framework that naturally complements economic theory and supports many implementations of bounded rationality, i.e., training boundedly rational RL agents. 
In this work, we showed an instance of this idea, by modeling the effect of inattention through imposing costs on observing information. 
Our analysis demonstrates that multi-agent RL can yield new insights into the effects of bounded rationality.

Moving forward, multi-agent RL could assist in accelerating the understanding of bounded rationality.
For instance, one could collect observational behavioral data to ground and justify more sophisticated boundedly rational agents and their reward functions.
RL agents trained on these novel reward functions could then produce falsifiable predictions on the behavior of humans in unseen tasks and domains. 
Here, multi-agent RL could be incorporated in a continuous loop between data-collection and simulation-based analysis and prediction.

Future research could also look into the relationship of learning itself and bounded rationality. 
This relates to research questions around how humans acquire new information and how they use it to update their beliefs and ultimately their behaviors.

As such, we believe applications and extensions of reinforcement learning hold promise in advancing understanding bounded rationality and economic interactions at large, and are a highly promising direction for future work.

\paragraph{Limitations.} 
While RI is a general model of bounded rationality, it does not cover all models of human irrationality. 
For simulations that wish to consider more specialized models of bounded rationality, \ourframework{} may not be sufficient. 
Examples include hyperbolic discounting~\citep{kirby1995preference} to model time-inconsistent delay discounting (humans tend to prefer rewards in the near future much more than rewards farther in the future) or prospect theory~\citep{Tversky1992} to model loss-aversion.
Future research may extend \ourframework{} to include such notions of bounded rationality--whether towards improving understanding of Principal-Agent problems or applied more generally towards refining understandings of economic systems under more realistic assumptions of behavior.
\section{Ethics Statement}
%
%
Our work proposes a framework to model bounded rationality in Multi-Agent Reinforcement Learning based simulations.
Thus our framework may be used to draw implications in simulations modeled after real-world systems. 
While capturing human-like irrationalities in the behaviors of RL actors is an important step towards this goal, it is hardly sufficient for achieving the realism required for real-world decision making based on AI simulations.
We anticipate a much longer path forward before this technology matures to such a degree and note that any premature usage of our framework towards real-world decision making would contradict its intended research purpose.
Furthermore, our framework should not be used to explore methods to increase discrimination or unfairness in real-world systems; instead, it should be use to investigate how to decrease such biases. 
We acknowledge that our PA settings identify a tension between employer (Principal) profit and employee (Agent) utility/equality.
Our results are not meant to provide any actionable insight into how this tension could be manipulated for the benefit of one party.

%


\bibliography{references}
\bibliographystyle{iclr2021_conference}

\appendix

\section{Training Details and Hyperparameters}
\label{section:training-details-appendix}

\parwrapper{Pay Schedule Optimization in a Single-Step Environment.}
We used learning rates of 1e-3 for the training the principal policy parameters and 5e-3 for the mutual information classifier. We used a batch size of 128 and trained the principal for a total of $100000$ batches. During training we gradually annealed $\aplamb$ from $0$ to the desired value at a rate of $4/10000$ per batch. We average all results across 5 random seeds and we set the random seed for pytorch, numpy, and python's internal random module for each run. We used seeds of $[0,4]$. All experiments were run on 16CPU cloud compute machines with 54GB of memory. Given a pay schedule which consists of mean and standard deviation parameters $(\bm{\mu_\output}, \bm{\sigma_\output})$ for each output level $\output \in \Output$, to calculate the principal policy we first sample 100 pay schedules and calculate the agent utility per output for each output level for each pay schedule. We then average to calculate the average agent utility for each output level. We use the noise structure to calculate the average utility for each action and use the soft-q formulation over the utilities given in the main text to obtain the agent's stochastic policy.

\parwrapper{Wage Optimization in a Sequential, Multi-Agent Environment.}
We used learning rates of 1e-4 for the Principal and Agent's policy parameters and 1e-3 for all the mutual information classifiers. We used a batch size of 512 episodes and train the principal and agent through 60000 batches. We train a single \ourpolicy{} for the Agents and concatenate the experiences of all $\numa$ Agents when updating the policy. The Agent policy therefore effectively has a batch size of $512 \numa$. To avoid vastly different total episode returns, we scaled the rewards by the horizon during training. We run all experiments on 8CPU cloud computing machines with 26GB of memory. We average all results across 20 random seeds and we set the random seed for pytorch, numpy, and python's internal random module for each run. We used seeds of $[0,19]$.

\section{RIRL Implementation Details}
\label{section:implementation-details-appendix}

We use this section of the appendix to cover important implementation details and tips.
This is intended for practical guidance.

\paragraph{Stochastic Encoder Module Configuration and Initialization.}
Section~\ref{section:methods} describes the architecture of our \ourpolicy{} policy class.
As described, we learn an encoder $\encoder^\chn (y^\chn_t | \ob^\chn_t, \lstmstate_t)$ for each observation channel $\chn$.
Encoder $\encoder^\chn$ takes observation $\ob^\chn_t$ and recurrent state $\lstmstate_t$ as inputs and outputs the parameters (means and standard deviations) of a stochastic encoding $y^\chn_t$.

We find that, in practice, learning does not progress if each $y^\chn$ contains very little information about $\ob^\chn$ at the start of training.
To address this, we recommend two implementation choices.
First, implement $\encoder^\chn$ as a residual-style module:
\eq{
    \mu^\chn_t, \sigma^\chn_t = \encoder^\chn( \ob^\chn_t, \lstmstate_t ), \quad
    y^\chn_t = \ob^\chn_t + \mu^\chn_t + \sigma^\chn_t \cdot \epsilon^\chn_t,
}
This simply requires setting the output $y^\chn$ to have the same size as observation $\ob^\chn$ and adding $\ob^\chn_t$ to the mean $\mu^\chn_t$.
Second, initialize the output layer of $\encoder^\chn$ such that $\sigma^\chn$ is consistently very small.
We perform this by adding a constant negative offset to the bias units associated with the $\log \sigma^\chn$ outputs, which we exponentiate to get $\sigma^\chn$.
As a result of this strategy, $y^\chn_t$ closely follows $\ob^\chn_t$ at the start of training.

\paragraph{Hidden State as an Encoder Input.}
We emphasize that the inputs to encoder $\encoder^\chn$ is the concatenation of the observation $\ob^\chn_t$ and the hidden state $\lstmstate_t$.
Similarly, when using the discriminator $d^\chn(y^\chn_t, \left[\ob^\chn_t, \lstmstate_t \right])$ to estimate $\tilde{I}_{m}$ and when training the discriminator, we also apply this concatenation.
In other words, $d^\chn$ regards $\left[\ob^\chn_t, \lstmstate_t \right]$ as the observation, such that the $\tilde{I}_{m}$ captures the MI between $y^\chn_t$ and $\left[\ob^\chn_t, \lstmstate_t \right]$.

This is an important detail for ensuring that MI regularization works as expected in the multi-step setting.
For instance, we observed that, if the discriminator does not see the hidden state $\lstmstate$, $\encoder^\chn$ learns to encode $\ob^\chn$ such that $I(y^\chn; \ob^\chn)$ is minimal but where $\ob^\chn_t$ can still be easily recovered given $y^\chn_t$ \textit{and} $\lstmstate_t$.

\paragraph{Optimization}

During training, we found that learning was most stable if separate learning rates were used for the policy modules 
$\{ \encoder^1, \ldots, \encoder^\Chn, \texttt{LSTM}, \omega \}$
and for the discriminator modules
$\{ d^1, \ldots, d^\Chn, d^\omega \}$.
Importantly, \textit{the discriminator modules use a 10x higher learning rate.}
Concretely, we use learning rates of 0.0001 and 0.001 for the policy and discriminator modules, respectively.
Configuring learning rates this way helps to ensure that discriminator $d^\chn$ can adjust to changes in encoder $\encoder^\chn$ faster than the encoder can adapt to changes in the discriminator.
Intuitively, this improves the quality of the MI estimates during training.

Another important optimization detail concerns gradient flow.
Gradients from $\nabla \log \omega(a_t | \cdot)$ need to backpropagate through the encodings $\left[y^1_t, \ldots y^\Chn_t\right]$ in order for the encoder modules to receive meaningful gradients.
In Pytorch, which we use for this implementation, ensuring that this gradient flow occurs requires some attention.
Given the (learned) mean and standard deviation parameters (which are functions of the encoder input), Pytorch constructs the sampling distribution as
\texttt{m = Normal($\mu, \sigma$)}.
The \textit{output} of the encoder module $y^\chn_t$ must be sampled via the reparameterization trick:
\texttt{y = m.rsample()}, which allows gradients to flow through $y^\chn_t$.
Finally, this output should be detached from the backpropagation graph when calculating its \textit{log probability}:
\texttt{encoder\_log\_prob = m.log\_prob( y.detach() )},
which is needed for computing the policy gradients.
Similarly, care must be taken to ensure that inputs to the discriminators are detached in the same way.

\end{document}